\documentclass[prl,aps,twocolumn,superscriptaddress,nofootinbib]{revtex4-2}
\usepackage{inputenc}
\usepackage{epsfig}
\usepackage{amsfonts}
\usepackage{latexsym}
\usepackage{amsmath}
\usepackage{amssymb}
\usepackage{slashed}
\usepackage{longtable}
\usepackage{mathrsfs} 
\usepackage{bm,bbm}
\usepackage[normalem]{ulem}

\usepackage{graphicx}
\usepackage{hyperref}

\numberwithin{equation}{section}

\newcommand{\bea}{\begin{eqnarray}}
\newcommand{\eea}{\end{eqnarray}}
\newcommand{\be}{\begin{equation}}
\newcommand{\ee}{\end{equation}}
\newcommand{\ba}{\begin{align}}
\newcommand{\ea}{\end{align}}

\def\Or[#1]{{\text{O}}\left({#1}\right)}
\def\dotl[#1,#2]{\left\langle #1, #2 \right\rangle}
\def\dotlb[#1,#2]{[ #1, #2 ]}
\def\dotp[#1,#2]{(#1) \cdot (#2)}
\def\aff[#1,#2]{\hat{#1}(#2)}
\def\n4sym{{\cal N}=4 SYM}
\def\>{\rangle}
\def\<{\langle}
\def\weight[#1,#2,#3]{\{(#1),#2,#3\}}
\def\ads[#1]{$\text{AdS}_{#1}$}

  \makeatletter
  \let\over=\@@over \let\overwithdelims=\@@overwithdelims
  \let\atop=\@@atop \let\atopwithdelims=\@@atopwithdelims
  \let\above=\@@above \let\abovewithdelims=\@@abovewithdelims

\usepackage{xcolor}

\begin{document}
\title{Dual gauge theory formulation of planar quasicrystal elasticity and fractons}

\author{Piotr Sur\'{o}wka}
\email{surowka@pks.mpg.de}
\affiliation{Department of Theoretical Physics, Wroc\l{}aw  University  of  Science  and  Technology,  50-370  Wroc\l{}aw,  Poland}
\affiliation{Max Planck Institute for the Physics of Complex Systems and W\"{u}rzburg-Dresden Cluster of Excellence ct.qmat, 01187 Dresden, Germany}

\begin{abstract}
Elastic description of planar quasicrystals can be formulated as an interplay between two Goldstone fields corresponding to phonon and phason degrees of freedom. We reformulate this description as a gauge theory with one gauge field that is symmetric under exchange of indices and one that is not. We also show that topological defects in quasicrystals can be succinctly incorporated in the dual description and interpret them as fractonic excitations. Finally we calculate the static interaction potential between defects in a quasicrystal with fivefold symmetry. This is done in the limit of a small coupling between phonon and phason stresses.


\end{abstract}

\maketitle
 Solid materials are most commonly represented by crystals, whose atomic constituents are arranged in a highly ordered microscopic structure, forming a lattice. Macroscopic description of crystals is provided by the theory of elasticity that deals with mechanics of bodies modelled as a continuous object rather than a crystalline lattice of atoms \cite{LandauLifshitz-7}. An important point is, however, that not all solids are crystals. In fact there are various classes of solid materials, whose microscopic structure is not a periodic lattice. Examples include policrystals, glasses and quasicrystals. We do not have such a detailed understanding of these materials as for crystals. Both of the microscopic structure and macroscopic descriptions are an active field of study.

Topological defects play a key role in the physics of elastic solids \cite{Nabarro}. They are characterized by a discontinuity in the order parameter. In the context of classical elasticity there are two types of such discontinuities: dislocations and disclinations. They are crucial in the two-dimensional phase transitions as first shown by Kosterlitz and Thouless and applied to solids by Nelson, Halperin and Young \cite{MichaelKosterlitz2013,NelsonHalperin1979,Young1979}. The transition happens through the proliferation of topological defects at finite temperature that results in a thermal melting of a crystal. In elasticity such melting occurs in two stages. First the dislocations condense, while the disclinations are still energetically too costly. This is the hexatic state, which is an example of the quantum nematic order \cite{Nelson}. In the next stage both dislocations and disclinations are condensing leading to the isotropic state. Although defects is a classic subject in the science of materials, the intricate geometric constructions are quite far from an effective field theory framework usually employed to study phase transition. To circumvent this issue a major theoretical insight, that allows one to easily incorporate defects as sources in the field theory formulation, is provided by Kleinert \cite{kleinert1982duality,kleinert1983double} (for a review see \cite{Kleinert,Kleinert1995,Beekman2017rev}). In a complete analogy to the particle-vortex duality he rewrites elasticity as a symmetric gauge field. This duality maps defects to matter fields charged by the dual gauge fields.

Symmetric tensor gauge fields emerge also as a low-energy description of certain spin liquids \cite{Xu2006,Xu2010,Pretko2017spinliquid,Pretko2017spinliquid2,You2020}. Such gauge fields are sourced by matter fields with restricted mobility, whose presence indicate a new type of topological phase of matter \cite{Nandkishorereview,Pretko:2020cko}. This similarity with quantum elasticity can be made precise in the language of dualities formulated earlier. Dislocations are vector charges that can move along their Burgers vector, while disclinations are fully immobile scalar charges. This behavior leads to the conclusion that elastic defects are in fact fractons \cite{Pretko2018}. As a result elastic dualities serve to expand our knowledge on the classical and quantum elasticity as well as to give insights into new fractonic phases of matter \cite{Zaanen2004,Cvetkovic2006,Beekman2017,PretkoSolid2018,Gromov2019elastic,Kumar2019,pretko2019crystal,zhai2019two,Gromov:2019waa,Nguyen:2020yve,Nampoothiri2020,Fruchart2020dual,Manoj:2020abe,Zhai2020}.

In this paper we intend to study the connection between fractons and elasticity of quasicrystals. Quasicrystals are solids, with long-range positional order and no periodicity \cite{levine1984quasicrystals,levine1985elasticity,socolar1986phonons,Baggioli:2020haa} (See \cite{Fan2016} for a review). This means that the elastic description of quasicrystals is fundamentally different than the one for crystals. The free energy of ordinary crystals is unchanged under a discrete translation corresponding to the lattice vectors defining the unit cell of the periodic structure. When the translation is allowed to vary slowly as a function of the position the free energy increases. This can be parameterized by a displacement field $u_i(x)$, which describes phonons, Goldstone bosons that are low-energy collective excitations of the crystal. In quasicrystals the free energy also remains constant under the global rearrangements of atomic positions that can be parameterized by an additional vector $w_i(x)$. Small fluctuations of this vector introduce the so-called phason field or strain, capturing the low-energy collective excitations of quasicrystal called phasons.

The main goal of the present work is to reformulate the elastodynamics of quasicrystals \cite{Ding1993} in terms of gauge theories. Such a formulation allows for a systematic study of defects and their interactions as well as it opens up a possibility to investigate two-dimensional defect-mediated phase transitions present in quasicrystals. It also provides a theoretical background for quantum phases with quasicrystalline symmetries.

{\it Elasticity of quasicrystals}.$-$Quasicrystals are characterized by two types of displacement fields $u_i(x)$ and $w_i(x)$. The first is analogous to the phonon displacement and leads to the symmetric strain tensor $u_{ij}=u_{ji}$, where $u_{ij}=\frac{1}{2}(\partial _i u_j+\partial_j u_i).$ Contrary to the phonon field $u_{ij}$ the phason displacement tensor $w_{ij}=\partial _i w_j$ is not symmetric $w_{ij}\neq w_{ji}$. Following the usual procedure of writing a free energy as an expansion around the zero displacement $u_{ij}=w_{ij}=0$ one can write down the potential part of the effective action $S=\int dt d^2x \mathcal{L}$ as
\begin{align}
\label{eq:freeen}
S_{\text{pot}} [u_i,w_i]&= \int dt d^2x \frac{-1}{2} \Big( C^{ijkl} u_{ij} u_{kl}\Big)\,\nonumber \\ 
& +\int dt d^2x\frac{-1}{2}  \Big( K^{ijkl} w_{ij} w_{kl}\Big)\\
&+\int dt d^2x \frac{-1}{2} \Big( R^{ijkl} w_{ij} u_{kl} + R'^{ijkl} w_{ij} u_{kl}\Big)\nonumber \\
& \equiv \int \frac{-1}{2}  \Big[ \left( u_{ij} \, w_{ij} \right) \begin{pmatrix} C_{ijkl} & R_{ijkl} \\R'_{ijkl}  & K_{ijkl}  \end{pmatrix} \begin{pmatrix}  u_{kl} \\ w_{kl} \end{pmatrix}  \Big],\, \nonumber
\end{align}
where we have introduced four tensors of elastic coefficients parameterizing different couplings between fluctuations. We always assume a summation over repeated indices, which run over spatial coordinates $x$ and $y$ as we focus on two-dimensional systems. It is important to note that the phonon elastic tensor $C_{ijkl}$ possesses both minor and major symmetries, the couplings between phonons and phasons represented $R^{ijkl}$ and $R'^{ijkl}$ have a minor symmetry of indices contracted with the phonon field. Finally the $K^{ijkl}$ has neither minor nor major symmetries. The  non-zero entries of elastic tensors in quasicrystals can be fixed by group theory in analogy to crystals. From a known action of discrete symmetries one determines quadratic invariants and associated elastic coefficients (see \cite{Hu_2000} for a review).  For two-dimensional quasicrystals such a procedure has been carried out for allowed fivefold \cite{levine1985elasticity}, eightfold and twelvefold symmetries \cite{Socolar1989}. Potential energy in the effective action can be supplemented by the kinetic part
\begin{equation}
S_{\text{kin}} [u_i,w_i] = \int dt d^2x \Big[\dot{u}_i \dot{u}_i  +\dot{w}_i \dot{w}_i\Big]\,.
\end{equation}
It is well known that the phason field is diffusive \cite{Lubensky1985,Francoual2003}, therefore the above elastodynamics describes either short time behavior of classical quasicrystals or systems at zero temperature. Several realizations of quantum systems with quasicrystalline symmetry have been proposed. These include bosons on optical lattices \cite{Gopalakrishnan2013,Kraus2013,Bandres2016} and correlated electron systems \cite{Tran2015,Sagi2016,Huang2018,Ahn2018,Varjas2019,Ochoa2019}. The partition function for quasicrystals elasticity reads
\begin{equation}
Z=\int Du^i Dw^i e^{iS_{\text{kin}} [u_i,w_i]+iS_{\text{pot}}[u^i, w^i]}\,.
\end{equation}
Our intention is to perform the duality transformation on the above action. In order to do that we need to write it first in terms of stress variables. There are two equivalent ways of doing that. One can directly perform the Hubbard-Stratonovich transformation or write down generalized Hooke's laws
\begin{subequations}
\begin{align}
\label{eq:stresses1}
T_{ij}&= -\frac{\partial \mathcal{L}}{\partial u_{ij}}=C_{ijkl}u_{kl}+R_{ijkl}w_{kl},\\
\label{eq:stresses2}
H_{ij}&= -\frac{\partial \mathcal{L}}{\partial w_{ij}}=K_{ijkl}w_{kl}+R'_{ijkl}u_{kl},
\end{align}
\end{subequations} 
solve for displacements in terms of stresses $T_{ij}$ and $H_{ij}$ and express the action in terms of these variables. To write down \eqref{eq:stresses1} and \eqref{eq:stresses2} we use the fact that $R'_{klij}=R_{ijkl}$. These transformations bring the action to the following form
\begin{align}
\label{eq:freeen2}
S [T_{ij},H_{ij},u_i,w_j]&= \int dt d^2x \frac{1}{2} \Big[ P_i P^i + \mathcal{P}_i \mathcal{P}^i\,\nonumber \\ 
& +\left( T_{ij} \, H_{ij} \right) \begin{pmatrix} C_{ijkl} & R_{ijkl} \\R'_{ijkl}  & K_{ijkl}  \end{pmatrix}^{-1} \begin{pmatrix}  T_{kl} \\ H_{kl} \end{pmatrix} \nonumber \\ 
&+ 2 u_i(\partial _\mu T^{i\mu})+2 w_i(\partial _\mu H^{i\mu})\Big], 
\end{align}
where we have introduced momentum operators $P^i=T^{i0}$ and $\mathcal{P}^i=H^{i0}$. Greek letters run over spacetime indices. After the transformation to stress variables displacements act as Lagrange multipliers for the conservation of momentum.
%

{\it Elastic duality for quasicrystals}.$-$Dualities offer new insights into non-perturbative physics of strongly correlated systems. In $2+1$ dimensions the core of dualities lies in the mapping of the Goldstone bosons fluctuations onto appropriate gauge fields. For example particle-vortex duality maps scalar fields onto $U(1)$ gauge fields and crystal elasticity maps strain fluctuations into symmetric tensor gauge fields. Elasticity of quasicrystals generalizes this mapping and introduces two sets of distinct elastic gauge fields. In order to see this we note that integrating out $u_i(x)$ and $w_i(x)$ we obtain two constraints coming from rewriting the action in terms of stress variables $\delta\left(\partial_\mu T^{i\mu}\right)$ and $\delta\left(\partial_\mu H^{i\mu}\right)$. In order to have a dual action for quasicrystals we resolve these constraints by two tensor gauge fields
\begin{equation}
T^{i\mu} = \epsilon^{\mu\nu\rho}\partial_\nu A^i_\rho\,,\qquad H^{i\mu} = \epsilon^{\mu\nu\rho}\partial_\nu \mathcal{A}^i_\rho\,.
\end{equation}
We note that a symmetric tensor such as $T^{i\mu}$ can be resolved by a symmetric tensor field $A_{ij}=A_{ji}$ and a scalar $\phi$ in a complete analogy with crystal elasticity
\begin{equation}
P^i = \epsilon^{kl}\partial_k A^i_l\,, \qquad T^{ij} = \epsilon^{jk}(-\partial_0A^i_k+\partial_k \partial _i \phi)\,,
\end{equation}
however, the resolution of the constraint for $H_{ij}$ leads to a tensor field $\mathcal{A}_{ij}\neq\mathcal{A}_{ji}$ that is not symmetric under exchange of indices and a vector potential $\Phi_i$
\begin{equation}
\mathcal{P}^i = \epsilon^{kl}\partial_k \mathcal{A}^i_l\,, \qquad H^{ij} = \epsilon^{jk}(-\partial_0 \mathcal{A}^i_k+\partial_k \Phi ^i)\,.
\end{equation}
In analogy with Maxwell electrodynamics we can define electric and magnetic fields
\begin{equation}
B^i =\epsilon^{kl}\partial_k A^i_l\,, \qquad E^i_j = \epsilon^i{}_k(-\partial_0A^k_j+\partial_j \partial _k \phi)\,.
\end{equation}
\begin{equation}
\mathcal{B}^i =\epsilon^{kl}\partial_k \mathcal{A}^i_l\,, \qquad \mathcal{E}^i_j = \epsilon^i{}_k(-\partial_0 \mathcal{A}^k_j+\partial_j\Phi^k)\,.
\end{equation}
These fields are invariant under the following gauge transformations
\begin{equation} \label{eq:gauge1}
\delta A_{ij} = \partial_i \partial_j \alpha\,, \qquad \delta \phi = \dot{\alpha}\,,
\end{equation}
\begin{equation}
\delta \mathcal{A}_{ij}= \partial_j \beta_i \, , \qquad \delta\Phi_i = \dot{\beta}_i\,.
\end{equation}
Already at this stage we see that the gauge structure for the symmetric stress tensor is the same as in scalar fracton theories and the asymmetric stress field leads to a vector gauge theory \cite{Pretko2018gaugeprinciple}. Thus quasicrystal elasticity combines these two degrees of freedom. We can now write the effective action in the dual formulation
\begin{align}
\label{eq:actiondual}
S_{\text{dual}}&= \int dt d^2x \frac{1}{2} \Bigg[ B_i B^i + \mathcal{B}_i \mathcal{B}^i\,\nonumber \\ 
& +\left( E_{ij} \, \mathcal{E}_{ij} \right) \begin{pmatrix} \tilde{\mathcal{C}}_{ijkl} & \tilde{\mathcal{R}}_{ijkl} \\ \tilde{\mathcal{R}}'_{ijkl}  & \tilde{\mathcal{K}}_{ijkl}  \end{pmatrix} \begin{pmatrix}  E_{kl} \\ \mathcal{E}_{kl} \end{pmatrix} \Bigg]\\ \nonumber 
&+\int dt d^2x\mathcal{L}_{\text{sources}}.\nonumber \\ \nonumber 
\end{align}
We have introduced a set of tensors that are yet to be fixed by calculating the proper inverse defined in \eqref{eq:freeen2}. We note that the entries are not just the inverses of the original elastic coefficients as we need to invert the whole matrix of tensors. Tilde denotes index rotations, e.g. $\tilde{\mathcal{C}}_{ijkl}= \epsilon_{ii'} \epsilon_{jj'}\epsilon_{kk'}\epsilon_{ll'}\mathcal{C}^{i'j'k'l'}$. The dual charges, that we later map to defects, couple to the gauge potentials in the following way
\begin{equation}
\mathcal{L}_{\text{sources}}=  \phi \rho +A_{ij}J_{ij}+ \Phi _i  \varrho_i+\mathcal{A}_{ij}\mathcal{J}_{ij} \,.
\end{equation}
The last ingredient that we need is the inverse matrix of elastic tensors. The explicit form of this matrix is in general complicated and depends on the symmetries of the quasicrystal that we want to study. In order to have a better intuition about the structure of this matrix w focus on a sub-class, for which the tensors of elastic coefficients can be decomposed into a set of projectors
\begin{subequations}
\begin{align}
\label{eq:decompc}
C_{ijkl}& = c_0 P^{(0)}_{ijkl}+c_1 P^{(1)}_{ijkl}+c_2 P^{(2)}_{ijkl},\\
K_{ijkl}& = k_0 P^{(0)}_{ijkl}+k_1 P^{(1)}_{ijkl}+k_2 P^{(2)}_{ijkl},\\
R_{ijkl}& = r_0 P^{(0)}_{ijkl}+r_1 P^{(1)}_{ijkl}+r_2 P^{(2)}_{ijkl},\\
R'_{ijkl}& = r'_0 P^{(0)}_{ijkl}+r'_1 P^{(1)}_{ijkl}+r'_2 P^{(2)}_{ijkl},
\end{align}
\end{subequations}
where the basis tensors are given by
\begin{subequations}
\begin{align}
\label{eq:project}
P^{(0)}_{ijkl}&= \frac{1}{2}\delta_{ij}\delta_{kl},\\
P^{(1)}_{ijkl}&= \frac{1}{2}(\delta_{ik}\delta_{jl}- \delta_{il}\delta_{jk}),\\
P^{(2)}_{ijkl}&= \frac{1}{2}(\delta_{ik}\delta_{jl}+ \delta_{il}\delta_{jk})-\frac{1}{2}\delta_{ij}\delta_{kl}.
\end{align}
\end{subequations}
One can check that $P^m_{ijab} P^n_{ab kl}=P^m_{ijkl}$ if $m=n$ and zero otherwise. Such a choice is of course an oversimplification, which will not apply to all systems. However, it illustrates the proof of concept behind this construction and the result can be presented in a compact form. In general, in order to construct the dual formulation for a specific quasicrystal, one has to specify elastic tensors and then explicitly construct an inverse. The simplest way to achieve this in two dimensions, for a given set of elastic parameters, is to pass to the Pauli matrix representation (see e.g. \cite{Scheibner2019,Banerjee2020,Fruchart2020dual}) and then apply an algorithm for the inversion of a block matrix \cite{LU2002119}. 

The inverse matrix is determined by the following matrix equation
\begin{equation}
 \begin{pmatrix} \mathcal{C}_{ijmn} & \mathcal{R}_{ijmn} \\ \mathcal{R}'_{ijmn}  & \mathcal{K}_{ijmn}  \end{pmatrix}\begin{pmatrix} C_{ijkl} & R_{ijkl} \\R'_{ijkl}  & K_{ijkl}  \end{pmatrix}=\begin{pmatrix} \text{Id}_{ijmn} & 0 \\0  &  \text{Id}_{ijmn} \end{pmatrix},
\end{equation}
where appropriate index contractions are understood after the multiplication of matrices. $\text{Id}_{ijmn}=\delta_{im}\delta_{jn}$ is the identity operator for four-tensors. For the case considered, this is a system of linear equations for twelve coefficients. The solution in the basis of projectors reads
\begin{subequations}
\begin{align}
\label{eq:decompcinv}
\mathcal{C}_{ijkl}& = \frac{k_0}{\Delta_0} P^{(0)}_{ijkl}+\frac{k_1}{\Delta_1}  P^{(1)}_{ijkl}+\frac{k_2}{\Delta_2}  P^{(2)}_{ijkl},\\
\mathcal{K}_{ijkl}& = \frac{c_0}{\Delta_0} P^{(0)}_{ijkl}+\frac{c_1}{\Delta_1} P^{(1)}_{ijkl}+\frac{c_2}{\Delta_2} P^{(2)}_{ijkl},\\
\mathcal{R}_{ijkl}& = - \frac{r_0}{\Delta_0} P^{(0)}_{ijkl}- \frac{r_1}{\Delta_1} P^{(1)}_{ijkl}- \frac{r_2}{\Delta_2} P^{(2)}_{ijkl},\\
\mathcal{R}'_{ijkl}& = -\frac{r'_0}{\Delta_0} P^{(0)}_{ijkl}- \frac{r'_1}{\Delta_1} P^{(1)}_{ijkl}-\frac{r'_2}{\Delta_2} P^{(2)}_{ijkl},
\end{align}
\end{subequations}
where $\Delta_m=c_m k_m+r_m r'_m$ for each coefficient labelled by $i\in \{0,1,2\}$. We note that it may happen the matrix of elastic coefficients is not invertible. In this case it is enough to construct the inverse in the invertible subspace as in the classical elasticity. Moreover, in analogy with block matrices not all individual entries have to be invertible for the existence of the inverse matrix. 

{\it Defects}.$-$Soon after the discovery of quasicrystals questions about the nature of defects, their dynamics and interactions became relevant \cite{levine1985elasticity,socolar1986phonons,LubenskyRamaswamyToner,Bohsung1987}. Several intricate geometric constructions are available, however, the intrinsic difficulty of these methods prevents us from having definite answers to all relevant questions about defects. Below we argue that the duality offers a natural, simple language for a systematic study of topological singularities in quasicrystals. In the dual language the elastic defects are mapped to the charges of the gauge theory. In order to see this mapping one can decompose the phonon and phason displacement fields into regular and singular parts $u^i = u^i_{\rm reg} + u^i_{\rm sing}$, $w^i = w^i_{\rm reg} + w^i_{\rm sing}$. Phonon displacement singularities couple to the conservation of the stress tensor
\begin{equation}
\delta S = \int dt d^2x \,\,\, \Big[u^i_{\rm sing}\partial_\mu T^{i\mu}\Big] =  \int dt d^2x \,\,\, \Big[\rho \phi+ J^{ij}A_{ij}\Big]\,.
\end{equation}
where $\rho=\partial^i \rho_i$, $\rho^i = \epsilon^{i}{}_j\epsilon^{kl}\partial_k\partial_l u^{j}_{\rm sing}$ and $J^{ij} =\epsilon^{i}{}_{n} \epsilon^{\mu\nu j}\partial_\mu\partial_\nu u^n_{\rm sing}$. In the last equality we employ an integration by parts. The charge $\rho$ is mapped to the disclination density
\begin{equation}
\rho_{\rm disc} =  \frac{1}{2} \epsilon^k{}_{l}  \epsilon^{ij}\partial_i\partial_j \partial_k u^l_{\rm sing}\,.
\end{equation}
We now study the coupling of the phason singularities $w^i_{\rm{sing}}$. Following the same logic as above we find
\begin{equation}
\delta S = \int dt d^2x \,\,\, \Big[w^i_{\rm sing}\partial_\mu H^{i\mu}\Big] =  \int dt d^2x \,\,\, \Big[\varrho_i \Phi^i+ \mathcal{J}^{ij}\mathcal{A}_{ij}\Big]\,,
\end{equation}
where we have introduced matching or stacking faults as the phason defects are sometimes dubbed in the literature
\begin{equation}
\varrho^i = \epsilon^{i}{}_j\epsilon^{kl}\partial_k\partial_l w^{j}_{\rm sing}
\end{equation}
and the current
\begin{equation}
\mathcal{J}^{ij} =\epsilon^{i}{}_{n} \epsilon^{\mu\nu j}\partial_\mu\partial_\nu w^n_{\rm sing}.\,
\end{equation}
We can now identify the duality mapping between charges and defects for phasons. Vector charges in the dual theory map to the rotated matching faults $ \epsilon^{i}{}_j \varrho^j$. With these mappings we can study the Gauss laws in the theory. They follow from the gauge transformations \label{eq:gauge1} in the Hamiltonian formulation of the theory
 \begin{equation}
\mathcal{H}= \Pi _{ij} \dot{A}_{ij} +\Pi^H_{ij} \dot{\mathcal{A}}_{ij} -\mathcal{L},
\end{equation}
where $\Pi _{ij}=\frac{\delta S}{\delta \dot{A}_{ij}} $ and $\Pi^H_{ij}=\frac{\delta S}{\delta \dot{\mathcal{A}}_{ij}} $ are the canonical momenta of the phonon and phason fields respectively. Invariance with respect to the gauge transformations leads to the following generalized Gauss laws
 \begin{equation}
\partial_{i} \partial _j \Pi^{ij}=\rho,
\end{equation}
 \begin{equation}
\partial_{i} \Pi^{ij}_H=\varrho^j.
\end{equation}
We see that these Gauss laws are exactly the same as the ones constructed for scalar and vector gauge theories \cite{Pretko2018gaugeprinciple}. As a result we can now identify defects in quasicrystals as different types of fractons. As far as the phonon field is concerned the defects are the same as in classical elasticity. We have disclinations that are immobile and dislocations corresponding to the disinclination dipoles that can move only along their Burgers vector. In addition to that the matching faults correspond to vector charges. The matching faults conserve the dipole moment $D_i$, where $D_i=\int x^i \partial _j \varrho ^j=0$. Therefore the matching faults are fractons with restricted mobility. Given this quasicrystals offer a natural playground to investigate scalar and vector fracton theories.

{\it Defect potential}.$-$We now apply the duality to study the static defect potential in a simple illustrative example of quasicrystals with fivefold symmetry in the limit of negligible phonon-phason coupling. The relevant part of the action reads
\begin{align}
\label{eq:freeen}
S&=  \int dt d^2x \frac{1}{2} \Bigg[  \left( \partial _i \partial _j \phi \, \partial _i \Phi_j \right) \begin{pmatrix} \tilde{\mathcal{C}}_{ijkl} & 0\\ 0  & \tilde{\mathcal{K}}_{ijkl}  \end{pmatrix} \begin{pmatrix}  \partial _k \partial _l \phi  \\  \partial _k \Phi_l  \end{pmatrix} \\ \nonumber
&+ \phi \rho + \Phi _i  \varrho_i  \Big].\, \nonumber
\end{align}
In the case of planar fivefold symmetry the elastic tensors are given by \cite{Ding1993}
\begin{subequations}
\begin{align}
\label{eq:decompcinv}
C_{ijkl}& = \lambda \delta_{ij}\delta_{kl}+\mu(\delta_{ik}\delta_{jl}+\delta_{il}\delta_{jk}),\\
K_{ijkl}& = K_1 \delta_{ik}\delta_{jl}+K_2(\delta_{ij}\delta_{kl}-\delta_{il}\delta_{jk}).
\end{align}
\end{subequations}
The above tensors can be expanded in the basis of projectors upon the following identifications: $c_0=2(\lambda +\mu)$, $c_1=0$, $c_2=2\mu$, $k_1=K_1+K_2$, $k_2=K_1+K_2$, $k_3=K_1-K_2$. $\lambda$ and $\mu$ are the first and second Lam\'{e} parameters. Integrating out the gauge fields $\phi$ and $\Phi_i$ one obtains the following expression,
\begin{equation}
\mathcal{L}=-\frac{1}{2} \rho_{\text{vec}}^{\text{T}}(-\mathbf{q}) \mathcal{V}  \rho_{\text{vec}}(\mathbf{q}),
 \end{equation}
where $ \rho_{\text{vec}}^{\text{T}} =\left( \rho \,\, \varrho_1 \,\, \varrho_2 \right)$. It gives the static potential between the defects. The disclination potential is $\mathcal{V}_{\rho \rho}= \frac{4\mu (\lambda+\mu)}{q^4(\lambda +2\mu)} $ and the explicit form of the matching fault potential is given by
\begin{align}
&\mathcal{V_{\varrho \varrho}}=  \nonumber \\
&\begin{pmatrix}  \frac{(K_1-K_2)\left[(K_1+K_2)q_1^2+2K_1 q_2^2\right]}{q^4K_1}   & -\frac{(K_1-K_2)^2q_1 q_2}{ q^4 K_1} \\   -\frac{(K_1-K_2)^2q_1 q_2}{q^4 K_1}  &     \frac{(K_1-K_2)\left[2 K_1q_1^2+(K_1+K_2) q_2^2\right]}{q^4K_1}    \end{pmatrix} .
 \end{align}
In the limit we consider there is no potential between matching faults and defects in the phonon field. Our computation shows the power of the duality that greatly simplifies the analysis by matching the defects into charges of appropriate gauge fields that can be easily dealt with using field theory methods.

{\it Discussion}.$-$We have constructed a dual formulation of quasicrystal elasticity. It has two stress tensors, one symmetric and one not, which ultimately leads to the dual description in terms of the two tensor gauge fields. Both these fields couple to fractonic charges dual to dislocations and disclinations in the phonon sector and to matching faults for phasons. It follows that quasicrystal elasticity is a natural place where scalar and vector fracton theories coexist. 

The duality offers a way to address open questions about the phase structure of quasicrystals. It was speculated in the past that a brittle-ductile transition can be related to the BKT-type of transition \cite{Kleman2003}. The dual formulation simplifies an analysis of phase transitions due to defect proliferation. Thus the construction presented here can be used as a starting point for a detailed study.

{\it Acknowledgements}.$-$PS was supported by the Deutsche Forschungsgemeinschaft through the Leibniz Program, the cluster of excellence ct.qmat (EXC 2147, project-id 390858490) and the National Science
Centre (NCN) Sonata Bis grant 2019/34/E/ST3/00405. 

\bibliography{Bibliography.bib}

\end{document}